\documentclass[12pt]{iopart}
\usepackage{dsfont}
\usepackage{amssymb}
\usepackage{graphicx}
\begin{document}

\title{Equation of motion for multiqubit entanglement in multiple independent noisy channels}
\author{Zhong-Xiao Man$^{1}$,Yun-Jie Xia$^{1}$ and Shao-Ming Fei$^{2,3}$}
\address{$^{1}$Shandong Provincial Key Laboratory of
Laser Polarization and Information Technology, Department of
Physics, Qufu Normal University, Qufu 273165, China}
\address{$^{2}$School of Mathematical Sciences, Capital Normal University,Beijing 100048, China}
\address{$^{3}$Max-Planck-Institute for Mathematics in the Sciences, 04103 Leipzig, Germany
\vskip0.5cm
E-mail: manzhongxiao@163.com; yjxia@mail.qfnu.edu.cn; feishm@mail.cnu.edu.cn}
\begin{abstract}
We investigate the possibility and conditions to factorize the entanglement evolution
of a multiqubit system passing through multi-sided noisy channels.
By means of a lower bound of concurrence (LBC)
as entanglement measure, we derive an explicit formula of
LBC evolution of the $N$-qubit generalized Greenberger-Horne-Zeilinger (GGHZ)
state under some typical noisy channels, based on which two kinds of factorizing
conditions for the LBC evolution are presented.
In this case, the time-dependent LBC can be
determined by a product of initial LBC of the system and the LBC evolution of a maximally
entangled GGHZ state under the same multi-sided noisy channels.
We analyze the realistic situations where these two kinds of factorizing conditions
can be satisfied. In addition, we also discuss the dependence of entanglement robustness on
the number of the qubits and that of the noisy channels.
\end{abstract}

\maketitle

\section {Introduction}
Quantum entanglement, as a type of distributed or nonlocal coherence
among several quantum subsystems, have been more and more recognized
as an indispensable resource in realizing many intriguing
quantum information processing and quantum computation \cite{QI}.
Realistically, however, the unavoidable coupling of
the entangled system to environment will lead to a destruction of the
necessary entanglement.
Therefore, a deeper understanding
of entanglement dynamics is of great importance not only in the foundation of quantum mechanics \cite{deep}
but also in the rapidly developing quantum technologies \cite{QI}.

The usual way in studying entanglement evolution is first to
deduce the evolved state
of the system and then to calculate its
entanglement \cite{dy1,dy2,dy3,dy4,dy5,dy6,dy7,dy8}.
However, as far as multipartite systems (or higher dimensional
systems) be concerned, it is generally very hard to solve the state equation
and therefore the evolution of entanglement can be determined
only in very special cases. Different from the aforementioned state-evolution technique,
it was found \cite{Konrad} for any two-qubit system
with only one qubit being subjected to a noisy environment, i.e., under one-sided noisy channel,
the evolution of system's entanglement in terms of
concurrence \cite{con} can be completely determined
by the product of the system's initial concurrence and the concurrence evolution of
a maximally entangled state under also the one-sided channel.
Through this channel-dependent technique one can
characterize the entanglement dynamics under unknown
channels by probing the entanglement evolution of a maximally entangled state alone
without exploring the concrete action of the channel on all initial states \cite{Konrad}.
The entanglement factorization law \cite{Konrad} has been experimentally verified
in an amplitude decay channel \cite{exper1}, and in the combined channel of phase damping and
amplitude decay \cite{exper2}.
The factorization law has been generalized to
finite-dimensional bipartite system for both initially pure state\cite{pure}
and mixed state\cite{mixed1,mixed2}, and to determine
the evolution of the G concurrence \cite{G} of two qudits \cite{qudit}.
In Ref.\cite{GG}, the author studied the multipartite system with
only one subsystem undergoing an arbitrary physical process and found that
the entanglement evolution of the multipartite system can be determined
uniquely by a single function of the quantum channel alone,
irrespective to the number of qudits in the system.
Therefore, in this case, the evolution of multipartite entanglement is in exactly the same way as
that of bipartite (two qudits) entanglement \cite{GG}.

In many practical situations, each particle of an entangled system
is coupled with a local noisy environment, therefore
the factorization law of entanglement evolution under one-sided noisy channel
should be generalized to the case of multi-sided channels.
It would be of great interest if the entanglement evolution
under multi-sided noisy channels could be factorized as that in the one-sided channel.
Unfortunately, however, even for the simplest case of two-qubit entanglement under
two-sided noisy channels, the factorization law \cite{Konrad}
does not hold any more except for some special forms of two-qubit state under
special environments \cite{Li1,Li2}.
It was shown that \cite{Li1} for the two-sided amplitude damping channels, only for the
NOE (the number of excitations in the system
is not more than one) state the factorization law is valid.
For the two-sided pure dephasing channel case, the factorization law is
valid not only for the NOE state but also for all the three basis states \cite{Li2}.
Therefore, a general factorization law of entanglement evolution
does not hold for multi-sided channels. However, it is still useful to explore the
factorization law for special forms of entangled states,
especially the multipartite entanglement, under special noisy channels. This study may
shed some light on the understanding of entanglement dynamics of multipartite system.
In this work, focusing on the
$N$-qubit generalized Greenberger-Horne-Zeilinger (GGHZ) state \cite{ghz},
we shall investigate the possibility and conditions to factorize the entanglement
evolution of multiqubit system under multi-sided noisy channels.

In studying the dynamics of multipartite entanglement, one of the biggest obstacles
is the lack of a computable entanglement measure.
Most studies \cite{GHZ1,GHZ2,GHZ3,GHZ4,MXA} on the entanglement dynamics of multiqubit
system are based on the strategy of bipartition and
the measure of bipartite entanglement. Another strategy one usually adopts is
to generalize the bipartite entanglement measure directly to the multipartite case \cite{GHZ5,GHZ6}.
However, the calculation for the mixed
state of a multipartite system requires an optimization process, which is very difficult to solve
exactly and can only be determined purely algebraically in the regime where the mixing is
moderate \cite{GHZ5}. To be able to assess the global entanglement of
a multiqubit system, we adopt the concept of lower bound of concurrence (LBC)\cite{LBC1,LBC}.
Based on the derived formula of LBC evolution of a $N$-qubit system under some typical noisy channels,
we find two kinds of sufficient conditions under which the factorization law
holds, namely, the time-dependent LBC can be
determined by the product of initial LBC of the system and the LBC evolution of a maximally entangled
$N$-qubit state under the same multi-sided noisy channels.
Then we consider the realistic situations under which the two kinds of factorization conditions can be satisfied.
The dependence of entanglement robustness on
the number of qubits and that of the noisy channels is also discussed.

\section{LBC evolution and the conditions for its factorization}
\subsection{Time-dependent LBC of a $N$-qubit system under multi-sided noisy channels}
In this work, we adopt the
LBC proposed in \cite{LBC} to quantify the global entanglement of a multiqubit system,
defined for an arbitrary $N$-qubit mixed state $\rho$ as
\begin{equation}\label{LBC}
\mathcal{\underline{C}}_{N}(\rho)
=\sqrt{\frac{1}{2^{N-1}-1}\sum_{\mathcal{P}}\sum_{l_{1}=1}^{k_{1}}\sum_{l_{2}=1}^{k_{2}}
(\mathcal{C}^{\mathcal{P}[k|N-k]}_{l_{1}l_{2}}(\rho))^{2}},
\end{equation}
with
\begin{equation}\label{C}
\mathcal{C}^{\mathcal{P}[k|N-k]}_{l_{1}l_{2}}(\rho)=\max\{0,\sqrt{\lambda^{\mathcal{P}[k|N-k]}_{l_{1}l_{2},1}}-
\sum_{i>1}\sqrt{\lambda^{\mathcal{P}[k|N-k]}_{l_{1}l_{2},i}}\}.
\end{equation}
In Eqs. (\ref{LBC}) and (\ref{C}), $k|N-k$ stands for the bipartitions with $k$ qubits in
one block and $N-k$ ones in another. In view of the distinctions of the $N$ qubits, we use $\mathcal{P}[k|N-k]$
to specify a concrete combination of $k$ and $N-k$ qubits in constituting the bipartitions
$k|N-k$. Thus, $\sum_{\mathcal{P}}$ stands for the summation over all possible
concrete bipartitions of $\mathcal{P}[k|N-k]$. In Eq.(\ref{C}),
$\lambda^{\mathcal{P}[k|N-k]}_{l_{1}l_{2},i}$ are the eigenvalues, in decreasing order, of
the non-Hermitian matrix $\rho(L_{l_{1}}^{k}\otimes L_{l_{2}}^{N-k})\rho^{*}
(L_{l_{1}}^{k}\otimes L_{l_{2}}^{N-k})$ with $\{L_{l_{1}}^{k};l_{1}=1,2...k_{1}\}$
and $\{L_{l_{2}}^{N-k};l_{2}=1,2...k_{2}\}$ the generators of group
SO($2^{k}$) and SO($2^{N-k}$) acting on the $k$ and $N-k$ qubits
of a concrete bipartition $\mathcal{P}[k|N-k]$.
Surely, $\mathcal{\underline{C}}_{N}(\rho)>0$ signifies
that $\rho$ is entangled and a separable state $\rho$
always has $\mathcal{\underline{C}}_{N}(\rho)=0$. Yet,
$\mathcal{\underline{C}}_{N}(\rho)=0$ does not necessarily imply separability of
$\rho$. The LBC of three-qubit X states was analyzed in Ref. \cite{LBCX}
to demonstrate when it goes to zero.

It is known the forms of multiqubit entanglement are diverse. In this work,
the consideration is restricted to the
GGHZ state of $N$ qubits in the form
\begin{equation}\label{GGHZ}
\left|\Psi_0\right\rangle_{1,2...N}=\alpha\left|i_{1},i_{2},...i_{N}\right\rangle_{1,2...N}
+\beta\left|\overline{i}_{1},\overline{i}_{2},...\overline{i}_{N}\right\rangle_{1,2...N},
\end{equation}
where $\alpha,\beta\in \mathbb{C}$ satisfying $|\alpha|^{2}+|\beta|^{2}=1$, $i_{1},i_{2}...i_{N}\in\{0,1\}$ and
$\overline{i}_{1}=1\oplus i_{1},\overline{i}_{2}=1\oplus i_{2}...\overline{i}_{N}
=1\oplus i_{N}$ with $\oplus$ being an addition mod 2.
Consider the $N$ qubits are locally subjected to $M\leq N$ independent noisy
channels without mutual interactions between the qubits. The dynamics of the $j$-th
qubit is governed by a master equation that gives rise to a completely
positive trace-preserving map (or channel) $\mathcal{E}_{j}$ describing
the evolution as $\rho_{j}(t)=\mathcal{E}_{j} \rho_{j}(0)$, where $\rho_{j}(0)$
and $\rho_{j}(t)$ are,
respectively, the initial and evolved states of the
$j$-th qubit. Under the actions of $M$ independent noisy channels,
the density matrix of the initial
state (\ref{GGHZ}) of the $N$-qubit system
\begin{eqnarray}\label{GGHZrho}
\rho_{N}(0)&\equiv&\left|\Psi_0\right\rangle_{1,2...N}
\left\langle\Psi_{0}\right|\nonumber\\
&=&|\alpha|^{2}
\left|i_{1},i_{2},...i_{N}\right\rangle_{1,2...N}\left\langle
i_{1},i_{2},...i_{N}\right|+|\beta|^{2}
\left|\overline{i}_{1},\overline{i}_{2},...\overline{i}_{N}\right\rangle_{1,2...N}\left\langle
\overline{i}_{1},\overline{i}_{2},...\overline{i}_{N}\right|\nonumber\\
&&+\alpha\beta^{*}\left|i_{1},i_{2},...i_{N}\right\rangle_{1,2...N}
\left\langle\overline{i}_{1},\overline{i}_{2},...\overline{i}_{N}\right|
+\alpha^{*}\beta\left|\overline{i}_{1},\overline{i}_{2},...\overline{i}_{N}\right\rangle_{1,2...N}\left\langle
i_{1},i_{2},...i_{N}\right|
\end{eqnarray}
will evolve into a
mixed state $\rho_{N}(t)$ given simply by the composition of the $M$
individual maps:
\begin{equation}
\rho_{N}(t)=\mathcal{E}_{1}\mathcal{E}_{2}...\mathcal{E}_{M}\rho_{N}(0).
\end{equation}
In this work, we shall consider several paradigmatic
types of noisy channels, such as the amplitude-damping (AD), depolarization (D)
and phase-damping (PD) (dephasing) channels.
Under the action of any one of these channels, it never appears new off-diagonal terms
in $\rho_{N}(t)$ but multiply the initial off-diagonal element of $\rho_{N}(0)$ (\ref{GGHZrho})
by a time-dependent factor,
whereas the diagonal elements of $\rho_{N}(0)$, may give rise to new diagonal terms \cite{GHZ4}. For convenience,
in the representation spanned by the $N$-qubit
product states $\{|00...00\rangle,|00...01\rangle,...,|11...11\rangle\}$,
we label the $2^{N}$ diagonal elements of $\rho_{N}(t)$
as $a_{1}(t),...,a_{m}(t),...,a_{2^{N-1}}(t)$,$b_{2^{N-1}}(t),...,b_{m}(t),...,b_{1}(t)$ (from the top left one to the
lower right one) and the two off-diagonal elements as $d_{m}(t)$ and $d_{m}^{*}(t)$.
Here, we specify the elements $a_{m}(t)$ and $b_{m}(t)$ ($d_{m}(t)$ and $d_{m}^{*}(t)$) as
the time-evolution of the initial nonzero diagonal (off-diagonal) elements of $\rho_{N}(0)$.
The other ones $a_{m^{\prime}}(t)$ and $b_{m^{\prime}}(t)$ (with $m^{\prime}\neq m\in\{1,2...,2^{N-1}\}$)
are the new diagonal elements derived in the time evolution.
Also, note that
the two diagonal terms of $\rho_{N}(t)$ corresponding to the elements $a_{i}(t)$ and $b_{i}(t)$ ($i=1,2...2^{N-1}$) have
completely opposite marginals for all the $N$ individual qubits.
As an example, suppose the initial density operator of a three-qubit system is
$\rho_{3}(0)=|\alpha|^{2}|000\rangle\langle000|+|\beta|^{2}|111\rangle\langle111|+\alpha\beta^{*}|000\rangle\langle111|
+\alpha^{*}\beta|111\rangle\langle000|$, the evolved density operator of which is $\rho_{3}(t)=
a_{1}(t)|000\rangle\langle000|+a_{2}(t)|001\rangle\langle001|+a_{3}(t)|010\rangle\langle010|+
a_{4}(t)|011\rangle\langle011|+b_{4}(t)|100\rangle\langle100|+b_{3}(t)|101\rangle\langle101|+
b_{2}(t)|110\rangle\langle110|+b_{1}(t)|111\rangle\langle111|+d_{1}(t)|000\rangle\langle111|
+d_{1}^{*}(t)|111\rangle\langle000|$. In this case, the time-evolution of the initial nonzero diagonal (off-diagonal)
elements are obviously $a_{1}(t)$ and $b_{1}(t)$ ($d_{1}(t)$ and $d_{1}^{*}(t)$).
The other diagonal elements $a_{m^{\prime}}(t)$ and $b_{m^{\prime}}(t)$ with $m^{\prime}=2,3,4$
are the new diagonal elements derived in the time evolution. The two diagonal terms, such as
$|001\rangle\langle001|$ and $|110\rangle\langle110|$ with the coefficients
$a_{2}(t)$ and $b_{2}(t)$ (the same for the other three pairs of diagonal terms with the coefficients
$a_{i}(t)$ and $b_{i}(t)$, $i=1,3,4$), involve completely opposite reduced states for all the three individual qubits.

By virtue of Eq. (\ref{C}), we obtain
the concurrence of a concrete bipartition $\mathcal{P}[k|N-k]$
of the evolved state $\rho_{N}(t)$ as
\begin{equation}\label{GHZC}
\mathcal{C}^{\mathcal{P}[k|N-k]}(\rho_{N}(t))=2\max\{0,|d_{m}(t)|
-\sqrt{a_{m^{\prime}}(t)b_{m^{\prime}}(t)}\},
\end{equation}
where $d_{m}(t)$ is the time-evolution of the initial off-diagonal element
of the GGHZ state (\ref{GGHZrho}), while $a_{m^{\prime}}(t)$ and
$b_{m^{\prime}}(t)$ (with $m^{\prime}\neq m$)
are the diagonal elements derived in the evolution.
The value of $m^{\prime} \in\{1,2...2^{N-1}\}$ is determined by the
definite bipartition $\mathcal{P}[k|N-k]$.
By virtue of Eqs. (\ref{LBC}) and (\ref{GHZC}), the LBC of the $N$-qubit system can be
expressed as
\begin{equation}\label{GHZLBC}
\mathcal{\underline{C}}_{N}(\rho_{N}(t))
=\sqrt{\frac{1}{2^{N-1}-1}\sum_{\mathcal{P}}
[\mathcal{C}^{\mathcal{P}[k|N-k]}(\rho_{N}(t))]^{2}}.
\end{equation}

So far, we have derived a general formula (\ref{GHZLBC}) of the time-dependent LBC of a $N$-qubit system under multi-sided noisy channels.
It should be pointed out that we have not solved the state evolution equation of the qubits' system
in the sense that we do not know the explicit forms of the matrix elements of the evolved density operator
$\rho_{N}(t)$. Actually, we are aware of the structure of the evolved density operator $\rho_{N}(t)$
since we are restricted to the GGHZ initial state of the $N$ qubits and to the special noisy channels,
which ensures the derivation of the formula (\ref{GHZLBC}).
Inspections of the formula (\ref{GHZLBC}) show that
to make the factorization law hold
for the LBC evolution of the $N$-qubit system under multi-sided noisy channels,
the associated bipartite concurrence $\mathcal{C}^{\mathcal{P}[k|N-k]}(\rho_{N}(t))$
(\ref{GHZC}) should satisfy either one of the following
two kinds of conditions.

\subsection{The first kind of condition}
An obvious situation under which the LBC (\ref{GHZLBC}) can be factorized is that
the product of $a_{m^{\prime}}(t)$ and $b_{m^{\prime}}(t)$ in the formula (\ref{GHZC})
is equal to zero,
i.e., $a_{m^{\prime}}(t)b_{m^{\prime}}(t)=0$ $\forall  m^{\prime}
\neq m\in\{1,2...,2^{N-1}\}$.
In this case, we have $\mathcal{C}^{\mathcal{P}[k|N-k]}(\rho_{N}(t))=2|d_{m}(t)|$
which is independent of the concrete bipartition $\mathcal{P}[k|N-k]$.
Accordingly, the LBC (\ref{GHZLBC}) of the $N$-qubit system is reduced to a simple form as
\begin{equation}
\mathcal{\underline{C}}_{N}(\rho_{N}(t))=
2|d_{m}(t)|,\label{LBCGHZ}
\end{equation}
in which $|d_{m}(t)|$ is obviously the time-evolution of the
off-diagonal element $|d_{m}(0)|=|\alpha\beta|$ of the $N$-qubit system.
That is, the LBC of the $N$-qubit system can be determined completely by the
time-evolution of the off-diagonal element $|d_{m}(t)|$.
As mentioned above, application of any one of these noisy channels to the $N$-qubit system
will multiply the off-diagonal element of $\rho_{N}(0)$ (\ref{GGHZrho})
by a time-dependent factor, namely,
$d_{m}(t)=d_{m}(0)\mathcal{D}(t)$.
Therefore, the LBC (\ref{LBCGHZ}) can be further reexpressed as
\begin{equation}\label{fac1}
\mathcal{\underline{C}}_{N}(\rho_{N}(t))=\mathcal{\underline{C}}_{N}(\rho_{N}(0))|\mathcal{D}(t)|
\end{equation}
where $\mathcal{\underline{C}}_{N}(\rho_{N}(0))=2|d_{m}(0)|=2|\alpha\beta|$ is the initial LBC
of the GGHZ state (\ref{GGHZ}). By virtue of Eq. (\ref{fac1}), for the initially maximal entangled state of the $N$-qubit
system with $\mathcal{\underline{C}}_{N}(\rho_{N}(0))=1$, we have
$\mathcal{\underline{C}}_{N}(\rho_{N}(t))=|\mathcal{D}(t)|$, which implies $|\mathcal{D}(t)|$ is
the LBC evolution of the maximally entangled state of the $N$-qubit system under the
same multi-sided noisy channels.
Therefore, the LBC evolution of the $N$-qubit system under multi-sided noisy channels
can be determined by the product of the initial LBC and the LBC evolution of
a maximally entangled state. Actually, as we shall show in next section, the factor $\mathcal{D}(t)$
can be determined by the parameters of the independent noisy channels that act on the qubits.
Of course, the factorization expression Eq. (\ref{fac1}) is conditioned on the condition
$a_{m^{\prime}}(t)b_{m^{\prime}}(t)=0$ in (\ref{GHZC}). In the next section, we shall show
that this condition can be satisfied in AD and PD channels.

\subsection{The second kind of condition}
In some situations, the first kind of condition $a_{m^{\prime}}(t)b_{m^{\prime}}(t)=0$ $\forall  m^{\prime}
\neq m\in\{1,2...,2^{N-1}\}$ may do not
hold and the LBC evolution cannot be factorized as the expression in Eq. (\ref{fac1}).
In this case, if $\sqrt{a_{m^{\prime}}(t)b_{m^{\prime}}(t)}$ can be
decomposed to $|d_{m}(0)||\mathcal{F}_{m^{\prime}}(t)|$,
the concurrence (\ref{GHZC}) of a definite bipartition $\mathcal{P}[k|N-k]$
will take the form
\begin{equation}\label{bicon}
\mathcal{C}^{\mathcal{P}[k|N-k]}(\rho_{N}(t))=
2|d_{m}(0)|\mathcal{Q}^{\mathcal{P}[k|N-k]}(t),
\end{equation}
with $\mathcal{Q}^{\mathcal{P}[k|N-k]}(t)=
\max\{0,|\mathcal{D}(t)|-|\mathcal{F}_{m^{\prime}}(t)|\}$.
By virtue of Eq. (\ref{GHZLBC}), LBC of the $N$-qubit system is thus reduced to
\begin{equation}\label{fac2}
\mathcal{\underline{C}}_{N}(\rho_{N}(t))=\mathcal{\underline{C}}_{N}(\rho_{N}(0))
\sqrt{\frac{1}{2^{N-1}-1}\sum_{\mathcal{P}}
(\mathcal{Q}^{\mathcal{P}[k|N-k]}(t))^{2}},
\end{equation}
with $\mathcal{\underline{C}}_{N}(\rho_{N}(0))=2|d_{m}(0)|
=2|\alpha\beta|$ being still the initial LBC of the GGHZ state (\ref{GGHZ}).
From Eq. (\ref{fac2}), for $\mathcal{\underline{C}}_{N}(\rho_{N}(0))=1$ corresponding to
an initially maximal entangled state of the $N$-qubit
system, we obtain its the LBC evolution as
$\mathcal{\underline{C}}_{N}(\rho_{N}(t))=\sqrt{\frac{1}{2^{N-1}-1}\sum_{\mathcal{P}}
(\mathcal{Q}^{\mathcal{P}[k|N-k]}(t))^{2}}$.
Therefore, the LBC evolution of an arbitrary $N$-qubit system under multi-sided noisy channels
can still be determined by the product of the initial LBC and the LBC evolution of
a maximally entangled state. In the next section, we shall show
that the LBC evolution of the $N$-qubit system under D channels can satisfy the above
condition and exhibit the factorized form in (\ref{fac2}).

\section{The factorization LBC evolution in various noisy channels}
We have shown theoretically that under certain conditions
the LBC evolution of a $N$-qubit system under multi-sided
noisy channels may be factorized in such a way that the LBC is determined by
the product of the initial LBC and the LBC evolution of a maximally entangled state under
the same multi-sided channels. Now, an immediate question arises that in what
realistic situations these conditions can be satisfied.
In the following, we provide an answer to this question by analyzing the evolution of
the $N$-qubit system in the AD, D and PD channels, respectively.

\subsection{Amplitude-damping channel}
At first, we consider the AD channel which correspond to the
zero-temperature dissipative reservoir.
The factorization law of two-qubit under one-sided noisy channel \cite{Konrad}
has been verified in the AD channel\cite{exper1,exper2}.
The action of an AD channel $j$
for a qubit is described by a map
$\mathcal{E}_{j}^{\mathrm{AD}}:\rho (t)=\mathcal{E}_{j}^{\mathrm{AD}}\rho (0)$ with
$\rho (0)$ and $\rho (t)$ the initial and evolved
density matrixes of the qubit.
During the time evolution a qubit decays from its excited
state $\left| 1\right\rangle $ to ground state $\left|
0\right\rangle $ by emitting an excitation, with a probability $p_{j}(t)=1-\exp
(-\Gamma_{j} t),$ where $\Gamma_{j} $ is the decay rate of the noisy channel.
The action of $\mathcal{E}_{j}^{\mathrm{AD}}$ on elements of the reduced density
matrix of a qubit reads:
\begin{equation}
\mathcal{E}_{j}^{\mathrm{AD}}:\left\{
\begin{array}{c}
\left| 0\right\rangle \left\langle 0\right| \rightarrow \left|
0\right\rangle \left\langle 0\right| , \\
\left| 0\right\rangle \left\langle 1\right| \rightarrow \sqrt{1-p_{j}(t)}%
\left| 0\right\rangle \left\langle 1\right| , \\
\left| 1\right\rangle \left\langle 0\right| \rightarrow \sqrt{1-p_{j}(t)}%
\left| 1\right\rangle \left\langle 0\right| , \\
\left| 1\right\rangle \left\langle 1\right| \rightarrow p_{j}(t)\left|
0\right\rangle \left\langle 0\right| +(1-p_{j}(t))\left| 1\right\rangle
\left\langle 1\right| .
\end{array}
\right.  \label{e0}
\end{equation}
Different forms of initial GGHZ state (\ref{GGHZ}),
such as $|\psi_{s}\rangle=\alpha|000\rangle+\beta|111\rangle$
and $|\psi_{a}\rangle=\alpha|001\rangle+\beta|110\rangle$ for a three-qubit system,
will exhibit different dynamical evolution in the AD channels \cite{MXA}.
Therefore, we classify the initial GGHZ state of the $N$-qubit system into symmetrical
and asymmetrical cases:
in the symmetrical GGHZ state (\ref{GGHZ1}) all the $N$ qubits are equivalent with the same reduced states,
while in the asymmetrical state (\ref{GGHZ2}) the $N$ qubits can be divided to two asymmetrical
blocks with opposite reduced states.

The symmetrical GGHZ state can be expressed as
\begin{equation}\label{GGHZ1}
\left|\Psi_\mathrm{I}\right\rangle_{1,2...N}=\alpha\left|0,0,...,0\right\rangle_{1,2...N}
+\beta\left|1,1,...1\right\rangle_{1,2...N}
\end{equation}
Without loss of generality, suppose the
qubits $1, 2,...,M$ ($M\leq N$) are locally coupled to $M$ independent and different AD channels, respectively.
Under the actions of the $M$ AD channels
the initial density matrix $\rho _{N}^{\mathrm{I}}(0)=\left| \Psi
_{\mathrm{I}}\right\rangle_{1,2...N} \left\langle \Psi _{\mathrm{I}}\right|$
will map onto $\rho_{N}^{\mathrm{I}}(t) $ which can be obtained as
\begin{eqnarray}\label{rhot1}
\rho_{N}^{\mathrm{I}}(t)&=&\prod_{j=1}^{M}[1-p_{j}(t)]^{\frac{1}{2}}
(\alpha\beta^{*}\left|00...0\right\rangle _{1,2...N}\left\langle 11...1\right|
+\alpha^{*}\beta\left|11...1\right\rangle _{1,2...N}\left\langle 00...0\right|)\nonumber\\
&&+|\beta|^{2}\sum_{i_{1}...i_{M}=0}^{1}q_{1}(t)...q_{M}(t)
\left|i_{1}...i_{M},1...1\right\rangle_{1...M,M+1...N}\left\langle i_{1}...i_{M},1...1\right|\nonumber\\
&&+|\alpha|^{2}\left|0...0\right\rangle_{1...N}\left\langle0...0\right|,
\end{eqnarray}
where the coefficients $q_{j}(t)\equiv p_{j}(t)$
$\forall i_{j}=0$ and $q_{j}(t)\equiv 1-p_{j}(t)$ $\forall i_{j}=1$ for $j=1,2,...,M$.
Recall that we specify $a_{m}(t)$ and $b_{m}(t)$ ($d_{m}(t)$ and $d_{m}^{*}(t)$)
as the time evolution of the nonzero diagonal (off-diagonal) elements of initial
density operator $\rho_{N}(0)$, while $a_{m^{\prime}}(t)$ and $b_{m^{\prime}}(t)$ (with $m^{\prime}\neq m$) are
the derived diagonal elements in the evolution.
From (\ref{rhot1}) we can see that the initial off-diagonal
element $|d_{m}(0)|=|\alpha\beta|$ has been evolved to
$|d_{m}(t)|=|\alpha\beta||\prod_{j=1}^{M}[1-p_{j}(t)]^{\frac{1}{2}}|$, while the initial
diagonal elements $a_{m}(0)=|\alpha|^{2}$ keep invariant and
$b_{m}(0)=|\beta|^{2}$ has been evolved to $b_{m}(t)=|\beta|^{2}(1-p_{1}(t))...(1-p_{M}(t))$.
To decide if the LBC of $\rho_{N}^{\mathrm{I}}(t)$ can be factorized, we should
check if the aforementioned condition $a_{m^{\prime}}(t)b_{m^{\prime}}(t)=0$ $\forall  m^{\prime}
\neq m\in\{1,2...,2^{N-1}\}$ can be satisfied.
From the second term of the right-hand side (RHS) of Eq. (\ref{rhot1}),
we observe that if not all the $N$ qubits are subjected
to the AD channels, i.e., $M<N$, all these diagonal terms have the same
reduced density operators $\left|11...1\right\rangle_{M+1,M+2...N}\left\langle 11...1\right|$
for the qubits $M+1,M+2...N$ that are not subjected to the noisy channels. In other words,
the matrix $\rho_{N}^{\mathrm{I}}(t)$ (\ref{rhot1}) does not involve the derived diagonal terms
whose reduced density operators for the qubits $M+1,M+2,...,N$ are
$\left|00...0\right\rangle_{M+1,M+2...N}\left\langle 00...0\right|$. Recall that the elements
$a_{m^{\prime}}(t)$ and $b_{m^{\prime}}(t)$ are the derived diagonal terms that should possess
opposite reduced density operators for all the $N$ individual qubits,
therefore in the case of $M<N$, we always have $a_{m^{\prime}}(t)b_{m^{\prime}}(t)=0$.
The LBC of the $N$-qubit system can thus be obtained by virtue of Eq. (\ref{fac1}) as
\begin{equation}\label{LBC1}
\mathcal{\underline{C}}_{N}(\rho^{\mathrm{I}}_{N}(t))=2|\alpha\beta|
\prod_{j=1}^{M}[1-p_{j}(t)]^{\frac{1}{2}}.
\end{equation}
Obviously, the $2|\alpha\beta|$ in (\ref{LBC1}) is the initial LBC of
the GGHZ state (\ref{GGHZ1}) while the term
$\prod_{j=1}^{M}(1-p_{j}(t))^{\frac{1}{2}}$ comprises the parameters of the $M$
AD channels.

In contrast to the case of $M<N$, if all the $N$ qubits
are coupled to the AD channels, i.e., $M=N$, then all the $2^{N}$ diagonal elements
of the states (\ref{rhot1}) are non-zero, which implies that the condition
$a_{m^{\prime}}(t)b_{m^{\prime}}(t)=0$ cannot be satisfied. Therefore,
the LBC of the system cannot be factorized to the form (\ref{LBC1}).
Actually, it has been shown \cite{GHZ4} under the action of $N$ independent AD channels,
the $N$-qubit system in the GGHZ state (\ref{GGHZ1}) will suffer from
the entanglement sudden death \cite{dy5} when $|\beta|>|\alpha|$.
At the same time, the robustness of the $N$-qubit system, i.e., the time at which the entanglement become
arbitrarily small, decrease with the increase of the number $N$ of qubits \cite{GHZ4}.
Here, for the case of $M<N$, we can see from the factorization expression (\ref{LBC1})
that the LBC is independent of the system size $N$ and only related to the
number $M$ of the AD channels.
In Fig. 1, we plot the evolution of LBC under $M=1,...,5$ identical AD channels
for the systems with $N>M$ qubits.
For convenience, we
have parameterized the time dependence in terms of $p\equiv p(t)$ instead of $t$
noticing that $p=0$ when $t=0$ and $p\rightarrow 1$ when
$t\rightarrow \infty ,$ i.e., $p\in [0,1]$ for $t\in [0,\infty ].$ From Fig.1,
we can see that the decay rate of the system increase with the number $M$ of the
AD channels. It deserves noting that for $M=1$, i.e., under one-sided channel,
the LBC evolution of any system with $N>2$
qubits are equivalent to the concurrence evolution of a two-qubit system under also the
one-sided channel, which is consistent with the result of Gour in \cite{GG}.

\begin{figure}[tbp]
\centerline{\scalebox{0.5}{\includegraphics{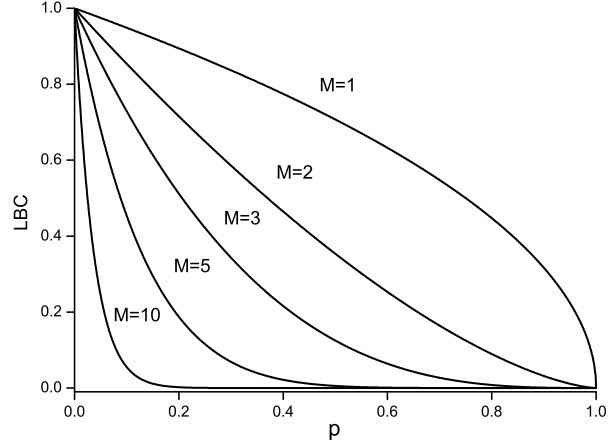}}}
\caption{The LBC evolution of a system with $N>M$ qubits that are initially
in the symmetrical GGHZ state (\ref{GGHZ1}) under actions of $M$ independent AD channels.
The parameter of the initial GGHZ state is $|\alpha|=|\beta |
=1/\protect\sqrt{2}$ and $p\in [0,1]$ for $t\in [0,\infty ]$.} \label{fig1}
\end{figure}

Next, we investigate the asymmetrical GGHZ state in which the reduced states of
the $N$ qubits are not equivalent.
Without loss of generality, we assume the reduced states of the qubits $1, 2,...,n$ with $1\leq n<N$ are all
$\left|0\right\rangle$, whilethe qubits
$n+1, n+2,...,N$ are all $\left|1\right\rangle$,
therefore the second class of GGHZ state can be expressed as
\begin{equation}\label{GGHZ2}
\left|\Psi_\mathrm{II}\right\rangle_{1,2...N}=\alpha\left|0...0,1...1\right\rangle_{1...n,n+1,...N}
+\beta\left|1...1,0...0\right\rangle_{1...n,n+1...N}.
\end{equation}
Suppose all the $N$ qubits are coupled with their own AD channels.
Then the initial density matrix $\rho _{N}^{\mathrm{II}}(0)=\left| \Psi
_{\mathrm{II}}\right\rangle_{1,2...N} \left\langle \Psi _{\mathrm{II}}\right|$
will map onto $\rho_{N}^{\mathrm{II}}(t) $ which can be obtained as
\begin{eqnarray}\label{rhot2}
\rho_{N}^{\mathrm{II}}(t)&=&\prod_{j=1}^{N}(1-p_{j}(t))^{\frac{1}{2}}
(\alpha\beta^{*}\left|0...0,1...1\right\rangle _{1...n,n+1...N}\left\langle 1...1,0...0\right|
+h.c.)\nonumber\\
&&+|\alpha|^{2}\sum_{i_{n+1}...i_{N}=0}^{1}q_{n+1}(t)...q_{N}(t)
\left|0...0,i_{n+1}...i_{N}\right\rangle_{1...n,n+1...N}\left\langle 0...0,i_{n+1}...i_{N}\right|\nonumber\\
&&+|\beta|^{2}\sum_{i_{1}...i_{n}=0}^{1}q_{1}(t)...q_{n}(t)
\left|i_{1}...i_{n},0...0\right\rangle_{1...n,n+1...N}\left\langle i_{1}...i_{n},0...0\right|,
\end{eqnarray}
where the coefficients $q_{j}(t)\equiv p_{j}(t)$
$\forall i_{j}=0$ and $q_{j}(t)\equiv 1-p_{j}(t)$ $\forall i_{j}=1$ with $j\in\{1,2,...,N\}$.
Here, we directly check if the first kind of condition $a_{m^{\prime}}(t)b_{m^{\prime}}(t)=0$ $\forall  m^{\prime}
\neq m\in\{1,2...,2^{N-1}\}$ can be satisfied by $\rho_{N}^{\mathrm{II}}(t)$.
From Eq. (\ref{rhot2}), we observe that there exist only two diagonal terms, i.e,
$\left|0...0,1...1\right\rangle_{1...n,n+1,...N}\left\langle0...0,1...1\right|$ and $
\left|1...1,0...0\right\rangle_{1...n,n+1...N}\left\langle1...1,0...0\right|$, which involve
completely opposite reduced states for all the $N$ individual qubits. However, these two terms
are merely the nonzero diagonal terms of the initial density operator $\rho _{N}^{\mathrm{II}}(0)$ rather than
the derived terms in the evolution. That is, the derived diagonal terms with elements
$a_{m^{\prime}}(t)$ and $b_{m^{\prime}}(t)$, such as $|0...0,1...0\rangle_{1...n,n+1,...N}\langle0...0,1...0|$
and $|1...1,0...1\rangle_{1...n,n+1,...N}\langle1...1,0...1|$, which involve
completely opposite reduced states for all the $N$ individual qubits will not appear simultaneously
(for the present example only the former term can appear while its counterpart does not).
Therefore, the condition $a_{m^{\prime}}(t)b_{m^{\prime}}(t)=0$ can
be satisfied for the $N$-qubit system in the GGHZ states (\ref{GGHZ2})
under the action of $N$ independent AD channels.
By virtue of (\ref{fac1}), we can get the factorized form of LBC for the
state $\rho_{N}^{\mathrm{II}}(t)$ as
\begin{equation}\label{LBC2}
\mathcal{\underline{C}}_{N}(\rho_{N}^{\mathrm{II}}(t))=2|\alpha\beta|\prod_{j=1}^{N}(1-p_{j}(t))^{\frac{1}{2}}.
\end{equation}
Obviously, $2|\alpha\beta|$ in (\ref{LBC2}) is the initial LBC of
the state (\ref{GGHZ2}) while the term
$\prod_{j=1}^{N}(1-p_{j}(t))^{\frac{1}{2}}$ reflects the actions of the $N$ AD channels.
Here we have considered all the $N$ qubits
are subjected to the independent AD channels, i.e., $M=N$. For the case of $M<N$, the factorized
form in (\ref{LBC2}) still hold but the RHS of (\ref{LBC2}) should be replaced by
$2|\alpha\beta|\prod_{j=1}^{M}(1-p_{j}(t))^{\frac{1}{2}}$
implying the action of $M$ AD channels.

\subsection{Depolarization channel}
The depolarizing channel $j$ describes the situation in
which a qubit remains untouched with probability
$1-p_{j}(t)$, or is depolarized (white noise), i.e., its
state is taken to the maximally mixed state, with probability
$p_{j}(t)$. The action of the map $\mathcal{E}_{j}^{D}$
of the D channel $j$ on elements of the reduced density
matrix of a qubit reads:
\begin{equation}
\mathcal{E}_{j}^{D}:\left\{
\begin{array}{l}
\left| 0\right\rangle \left\langle 0\right| \rightarrow (1-\frac{p_{j}(t)}{2})\left|
0\right\rangle \left\langle 0\right|+
\frac{p_{j}(t)}{2}\left|1\right\rangle \left\langle 1\right| , \\
\left| 1\right\rangle \left\langle 1\right| \rightarrow (1-\frac{p_{j}(t)}{2})\left|
1\right\rangle \left\langle 1\right|+
\frac{p_{j}(t)}{2}\left|0\right\rangle \left\langle 0\right|,\\
\left| 0\right\rangle \left\langle 1\right| \rightarrow (1-p_{j}(t))%
\left| 0\right\rangle \left\langle 1\right| , \\
\left| 1\right\rangle \left\langle 0\right| \rightarrow (1-p_{j}(t))%
\left| 1\right\rangle \left\langle 0\right|.
\end{array}
\right.  \label{e1}
\end{equation}
Without loss of generality, we suppose the
qubits $1, 2,...,M$ are coupled to $M$ ($M\leq N$) independent D channels, respectively.
Then under the actions of the $M$ different D channels
the initial density matrix $\rho _{N}(0)$ (\ref{GGHZrho}) of the $N$-qubit system
will map onto $\rho_{N}(t) $ which can be obtained as
\begin{eqnarray}\label{rhot3}
\rho_{N}^{D}(t)&&=\prod_{j=1}^{M}(1-p_{j}(t))
(\alpha\beta^{*}\left|i_{1}...i_{N}\right\rangle _{1...N}
\left\langle \overline{i}_{1}...\overline{i}_{N}\right|
+\alpha^{*}\beta\left|\overline{i}_{1}...\overline{i}_{N}\right\rangle _{1...N}
\left\langle i_{1}...i_{N} \right|)\nonumber\\
&&+|\alpha|^{2}\sum_{i_{1}^{\prime}...i_{M}^{\prime}=0}^{1}q_{1}(t)
...q_{M}(t)
\left|i_{1}^{\prime}...i_{M}^{\prime},i_{M+1}...i_{N}\right\rangle_{1...M,M+1...N}
\left\langle i_{1}^{\prime}...i_{M}^{\prime},i_{M+1}...i_{N}\right|
\nonumber\\
&&+|\beta|^{2}\sum_{\overline{i}_{1}^{\prime}...\overline{i}_{M}^{\prime}=0}^{1}g_{1}(t)
...g_{M}(t)\left|\overline{i}^{\prime}_{1}...\overline{i}^{\prime}_{M},\overline{i}_{M+1}...
\overline{i}_{N}\right\rangle_{1...M,M+1...N}
\left\langle \overline{i}^{\prime}_{1}...\overline{i}^{\prime}_{M},\overline{i}_{M+1}...\overline{i}_{N}\right|,
\end{eqnarray}
where the coefficients $q_{j}(t)\equiv1-\frac{p_{j}(t)}{2}$ $(\frac{p_{j}(t)}{2})$
when $i_{j}^{\prime}=i_{j}$ ($i_{j}^{\prime}=i_{j}\oplus1$),
$g_{j}(t)\equiv1-\frac{p_{j}(t)}{2}$ $(\frac{p_{j}(t)}{2})$ when $\overline{i}_{j}^{\prime}=\overline{i}_{j}$ ($\overline{i}_{j}^{\prime}=\overline{i}_{j}\oplus1$) with $j=1,2...M$.
Before discussing the general case, we first take a three-qubit system in the initial state $\rho_{3}(0)=|\alpha|^{2}|000\rangle_{123}\langle000|
+|\beta|^{2}|111\rangle\langle111|+\alpha\beta^{*}|000\rangle_{123}\langle111|+
\alpha^{*}\beta|111\rangle_{123}\langle000|$ as an example and suppose only the former two qubits
are subjected to the D channels. By virtue of Eq.(\ref{rhot3}), the evolved density operator read
\begin{eqnarray}\label{ex1}
\rho_{3}(t)&=&\prod_{j=1}^{2}(1-p_{j}(t))(\alpha\beta^{*}\left|000\right\rangle _{123}
\left\langle 111\right|
+\alpha^{*}\beta\left|111\right\rangle _{123}
\left\langle 000 \right|)\nonumber\\
&&+|\alpha|^{2}[(1-\frac{p_{1}(t)}{2})(1-\frac{p_{2}(t)}{2})|000\rangle\langle000|+\frac{p_{1}(t)}{2}\frac{p_{2}(t)}{2}|110\rangle\langle110|
\nonumber\\
&&+(1-\frac{p_{1}(t)}{2})\frac{p_{2}(t)}{2}|010\rangle\langle010|+\frac{p_{1}(t)}{2}(1-\frac{p_{2}(t)}{2})|100\rangle\langle100|]\nonumber\\
&&+|\beta|^{2}[(1-\frac{p_{1}(t)}{2})(1-\frac{p_{2}(t)}{2})|111\rangle\langle111|+\frac{p_{1}(t)}{2}\frac{p_{2}(t)}{2}|001\rangle\langle001|
\nonumber\\
&&+(1-\frac{p_{1}(t)}{2})\frac{p_{2}(t)}{2}|101\rangle\langle101|+\frac{p_{1}(t)}{2}(1-\frac{p_{2}(t)}{2})|011\rangle\langle011|].
\end{eqnarray}
From Eq. (\ref{ex1}), we can see that any one of the former four diagonal terms with a common coefficient $|\alpha|^{2}$
has a counterpart in group of the latter four terms with a common coefficient $|\beta|^{2}$,
such as $|110\rangle\langle110|$ and $|001\rangle\langle001|$, so as to these two terms have completely
opposite reduced states for all the three qubits. Therefore, the second factorization condition, i.e.,
$\sqrt{a_{m^{\prime}}(t)b_{m^{\prime}}(t)}$ can be decomposed to $|d_{m}(0)||\mathcal{F}_{m^{\prime}}(t)|$,
is satisfied for the present instance and we have
$\sqrt{a_{m^{\prime}}(t)b_{m^{\prime}}(t)}=|\alpha\beta||\mathcal{F}_{m^{\prime}}(t)|$ with
$\mathcal{F}_{m^{\prime}}(t)=\frac{p_{1}(t)}{2}\frac{p_{2}(t)}{2}$, $(1-\frac{p_{1}(t)}{2})\frac{p_{2}(t)}{2}$ or $\frac{p_{1}(t)}{2}(1-\frac{p_{2}(t)}{2})$
being determined by the concrete value of $m^{\prime}$. This specific example can be generalized
to the general case. Actually, for the evolved density operator (\ref{rhot3}), we observe that if $M<N$ the second
factorization condition hold. In this case, we have $a_{m^{\prime}}(t)=|\alpha|^{2}q_{1}(t)
...q_{M}(t)$, $b_{m^{\prime}}(t)=|\beta|^{2}g_{1}(t)
...g_{M}(t)$, and $\sqrt{a_{m^{\prime}}(t)b_{m^{\prime}}(t)}
=|\alpha\beta||\mathcal{F}_{m^{\prime}}(t)|$
with $|\mathcal{F}_{m^{\prime}}(t)|=\sqrt{q_{1}(t)
...q_{M}(t)g_{1}(t)...g_{M}(t)}$.
In addition, we note that the two diagonal elements $a_{m^{\prime}}(t)$ and $b_{m^{\prime}}(t)$ are equivalent so that
$|\mathcal{F}_{m^{\prime}}(t)|=|q_{1}(t)...q_{M}(t)|=|g_{1}(t)...g_{M}(t)|$.
By virtue of Eq. (\ref{fac2}), the LBC of the state $\rho_{N}^{D}(t)$ can be
expressed in the form
\begin{equation}\label{LBCD1}
\mathcal{\underline{C}}_{N}(\rho_{N}^{D}(t))=2|\alpha\beta|
\sqrt{\frac{1}{2^{N-1}-1}\sum_{\mathcal{P}}
(\mathcal{Q}^{\mathcal{P}[k|N-k]}(t))^{2}},
\end{equation}
where $2|\alpha\beta|$ is the initial LBC of
the GGHZ state (\ref{GGHZ}) and $\mathcal{Q}^{\mathcal{P}[k|N-k]}(t)
=\max\{0,|\prod_{j=1}^{M}(1-p_{j}(t))|-|\prod_{j=1}^{M}q_{j}(t)|\}$.
As for the case of $M=N$, the second kind of condition does not hold any more,
therefore the LBC evolution of the $N$-qubit system cannot be factorized as the
form (\ref{LBCD1}).

\begin{figure}[tbp]
\centerline{\scalebox{0.6}{\includegraphics{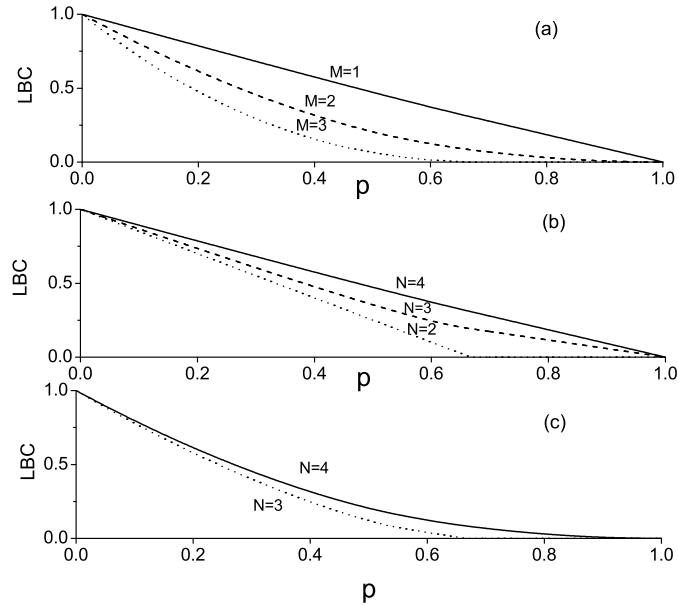}}}
\caption{(a) The time-evolution of LBC for a four-qubit system in the GGHZ state (\ref{GGHZ1})
under $M=1$ (solid line),
$M=2$ (dashed line) and $M=3$ (dotted line) D channels.
(b) The time-evolution of LBC for the system
with $N=4$ (solid line), $N=3$ (dashed line) and $N=2$ (dotted line) qubits under a single D channel.
(c) The time-evolution of LBC for the system
with $N=4$ (solid line), $N=3$ (dotted line) qubits under two D channels.
The parameters used are $|\alpha | =|\beta |
=1/\protect\sqrt{2}$ and $p\in [0,1]$ for $t\in [0,\infty ]$.} \label{fig2}
\end{figure}

In the following, we make a brief discussion on the relations between qubits' LBC
and the system's size $N$ as well as the number of the noisy channels $M$.
From (\ref{LBCD1}), we conclude that in the time-evolution of the $N$-qubit system,
the LBC increases with the qubit's number $N$
for the fixed number $M(<N)$ of the D channels,
while decreases with the number $M$ for the fixed qubit's number $N$.
In Fig.2 (a), we have shown the evolution of LBC of a four-qubit system
in the GGHZ state (\ref{GGHZ1}) under $M=1,2,3$ identical D channels,
where we can see the LBC decreases with an increase of $M$. Fig.2 (b) is for
the evolution of LBC for the systems
with $N=4$, $N=3$ and $N=2$ qubits under one-sided D channel, from which we can see
the robustness of LBC increases with $N$. In Fig.2(c) we plot
the evolution of LBC for the system
with $N=4$ and $N=3$ qubits under two identical D channels, which still
shows that the robustness of LBC increases with $N$.
Here, we have parameterized the time dependence in terms of $p\equiv p(t)$ instead of $t$
noticing that $p=0$ when $t=0$ and $p\rightarrow 1$ when
$t\rightarrow \infty ,$ i.e., $p\in [0,1]$ for $t\in [0,\infty ].$

\subsection{Dephasing channel}
The PD (or dephasing) channel
represents the situation in which there is loss of quantum
coherence with probability $p(t)$, but without any energy
exchange. The action of the map $\mathcal{E}_{j}^{PD}$
of the PD channel $j$ on the density matrix of a qubit multiplies the
off-diagonal elements by the factor $(1-p(t))$ while remaining the diagonal
elements invariant. Since the PD channels
cannot lead to the population evolution of the $N$-qubit system,
both the elements $a_{m^{\prime}}(t)$ and $b_{m^{\prime}}(t)$ representing the
derived diagonal terms in the evolution are zero. This implies that the first kind of condition for
the factorization of the LBC of the $N$-qubit system can always be satisfied
when the $N$-qubit system are coupled with independent PD channels. Therefore,
the LBC of the $N$-qubit system after evolution has the factorization form
in Eq. (\ref{fac1}).

\section{Conclusion}
In conclusion, by using LBC as a measure of multiqubit entanglement,
we have explored the possibility to generalize the factorization law of
two-qubit under one-sided channel to multiqubit under multi-sided channels.
Instead of a general answer to this issue, we have considered
a special form of multiqubit entanglement, i.e., the GGHZ state (\ref{GGHZ}),
and three typical types of noisy channels, i.e., the amplitude-damping (AD),
depolarizing (D) and phasing damping (PD) (dephasing) channels.
The explicit formulae (\ref{GHZLBC}) for the evolution of LBC as well as
the associated bipartite concurrence (\ref{GHZC}) for an arbitrary
bipartition are derived, based on which we observe that
under two kinds of conditions the LBC can be factorized.
That is, the LBC evolution of the $N$-qubit system can be determined
by the product of its initial LBC and the LBC evolution of a maximal
entangled state under the same multi-sided noisy channels.
We then study the realistic situations in which these two kinds of
conditions can be satisfied. We have shown that (i) the $N$-qubit
system in symmetrical GGHZ state (\ref{GGHZ1}) can satisfy the first
kind of factorization condition when the number $M$ of the independent
AD channels is lesser than the number $N$ of the qubits, while the $N$-qubit
system in asymmetrical GGHZ state (\ref{GGHZ2}) can always
satisfy the first kind of condition irrespective of the number $M$ of the
AD channels; (ii) in the D channels, the $N$-qubit system in the general
GGHZ state (\ref{GGHZ}) can satisfy the second factorization condition when
the number $M$ of the D channels is lesser than $N$; (iii) the PD channels can always lead to a factorization of LBC of
the $N$-qubit system in GGHZ state (\ref{GGHZ}) with a satisfaction of
the first kind of condition. By virtue of the concise
expressions of the LBC under different noisy channels, we have discussed
the dependence of the entanglement robustness on the system size in terms
of $N$ and the number $M$ of the noisy channels.
The GGHZ state, though only be a special class of various
types of multipartite entanglement, have attracted extensive researches from different aspects, such as
the preparation \cite{GHZpre}, application \cite{ghzapp},
dynamics\cite{GHZ1,GHZ2,GHZ3,GHZ4,GHZ5,MXA} and so on. Therefore, the study on
the factorization law of special multiqubit entanglement under some typical multi-sided
noisy channels can not only deepen the understanding of multipartite entanglement dynamics
but also facilitate the corresponding calculation.

\vskip 1cm

\noindent\textbf{Acknowledgments}

\noindent
This work was supported by National Natural Science Foundation of
China under Grant Nos.10947006 and 61178012, the
Specialized Research Fund for the Doctoral Program of Higher
Education under Grant No. 20093705110001, and the
Scientific Research Foundation of Qufu Normal University for Doctors.

\end{document}